# Test of the essential collapse-locality loophole.

Mónica Agüero, Juliana Bourdieu, Alejandro Hnilo, Marcelo Kovalsky and Myriam Nonaka.
*CEILAP, Centro de Investigaciones en Láseres y Aplicaciones, UNIDEF (MINDEF-CONICET);*
*CITEDEF,  J.B. de La Salle 4397, (1603) Villa Martelli, Argentina.*
email: alex.hnilo@gmail.com

March 25th, 2026.

Collapse-locality is an untested loophole in the violation of Bell's inequalities. The core of the argument is that the time value of photon detection is delayed by the time $T_c$ required by the collapse of its quantum state. The value of $T_c$ is given by the underlying theory of quantum collapse, and is mostly unknown. Depending on the value of $T_c$, detections in the performed Bell's experiments may have not been truly space-like separated events. This implies that the inequalities may have been violated as a consequence of (conspiratorial) information propagating at subluminal speed. We report an optical Bell experiment which closes the weaker ("essential") form of this loophole regardless the theory of quantum collapse. This is possible thanks to unique features of the setup. These features are: classical signals sent to the stations to define a time reference, and variable distance between the stations leaving all other parameters constant.

**Introduction.**

Local Realism (LR) names the intuitive belief that observed outcomes of an experiment are independent from events occurring outside the past light cone of said experiment, and that physical world's features are independent of being observed [1]. Some predictions of Quantum Mechanics (QM) (namely, violation of Bell's inequalities) seem incompatible with LR. A large number of experiments have been performed aimed to determine whether QM or LR is valid in the Nature. Practically all experiments have confirmed the violation of Bell's inequalities (BI), then disproving LR. Yet, experimental imperfections leave space to alternative LR theories to explain the observed results by exploiting the so-called *logical loopholes*. These alternative theories seem sometimes conspiratorial, but the consequences of giving up LR are so contrary to common sense that they are, at some extent, defensible. Closing the loopholes is therefore essential to the foundations of QM. As it was recently stated [2]: *"So it is fair to celebrate 2025 as the true centenary of quantum theory. Although such a commemoration can rightly point to a wide variety of breathtaking experimental successes, it must leave room to acknowledge the foundational questions that remain unanswered. QM is a beautiful castle, and it would be nice to be reassured that it is not built on sand"*.

The main loopholes can be roughly classified as follows [3]: *i*) *detection* loophole (or *fair sampling* or *no-enhancement*) which exploits imperfect efficiency of detection [4]; *ii*) *locality* (or *measurement independence*) loophole [5,6], which exploits that detection outcomes can be affected by signals from the other station propagating at subluminal speed, unless the settings are defined in an unpredictable and space-like separated way; *iii*) *time-coincidence* loophole, which exploits ambiguity in the definition of the time coincidence window [7]; *iv*) the *memory* loophole, which assumes detection outcomes are influenced by previous observations in such a way to fit the QM predictions [8].

Experiments performed during the last decade [9-16] practically closed all these loopholes. In some cases more than one loophole was closed simultaneously, no simultaneous closure of *all* known loopholes has been achieved. However, as it was stated in a critical review of the "loophole-free" experiments [17]: *"It should also be kept in mind that more loopholes may be found in the future... It is therefore conceivable that an experiment entirely free of loopholes cannot be done... What real experiments can do, in my opinion, is to shrink the space left to the loophole-based theories down to the point that (say, asymptotically) they become too exotic to be tenable. Also in my personal opinion, the reported experiments have already reached this point, and the defenders of LR should not insist on the little vulnerability left [by the known loopholes], but to pay attention to some barely explored alternatives"*.

In fact, a new loophole was found: the *collapse-locality* loophole [18]. Essentially, it indicates a weak point in the enforcement of the space-like separation of remote detections necessary to close the locality loophole. The weak point arises from the ignorance about the time taken by a quantum state to collapse (or to reduce, or to project). Depending on the actual value of this collapse time, the performed experiments may have left the locality loophole untested (see below).

The value of the collapse time is related with the open discussion on the nature of the quantum state, in short: whether it is physically real (*ontic* nature) or if it only represents the information the observer has about the system (*epistemic* nature) [19]. If the first alternative is true, it has been argued that the collapse time should take a finite value, as the changes of all physical entities do [20]. Several experiments to observe the collapse time, as it is predicted by existing theories of dynamical collapse, have been proposed [20-27]. In the second alternative, instead, the collapse is merely a change of the information accessible to the observer. It occurs in the observer's mind (as J. von Neumann argued). It is therefore not physical, and can be "instantaneous" (despite this feature is not relativistic covariant). Nevertheless, this alternative leaves the door open to the possibility that part of physical reality is left unaccounted by the quantum state (as Einstein argued) [28]. As the purpose here is to test the collapse-locality loophole, the ontic alternative is assumed in what follows.

The space-time diagram of a typical Bell's experiment with photons is displayed in the Figure 1.  In a naïf account, source S emits at time $T_e$ a pair of entangled photons that arrive to stations A and B at time $T_o$. It is usually assumed that, because of finite but short (and measurable) response time of instruments, photons are detected immediately after $T_o$. However [18]: *"…consider the possibility that the outcome of measurement A is actually determined by a collapse event at a point C in the causal future of the collapse event determining the outcome of measurement B… In this case the local hidden variables at C may depend explicitly on this prior collapse event, and hence on the choice [of setting] B and outcome b as well as on events at S"*.

There are two versions of the collapse-locality loophole. If the collapse event in C depends on the settings' choice in B (regardless a photon is actually detected there or not), then it is named *extended* loophole. If the event in C depends only on the outcome observed in B, then it is named *essential* loophole. In this paper, we report an experiment closing the *essential* loophole only.

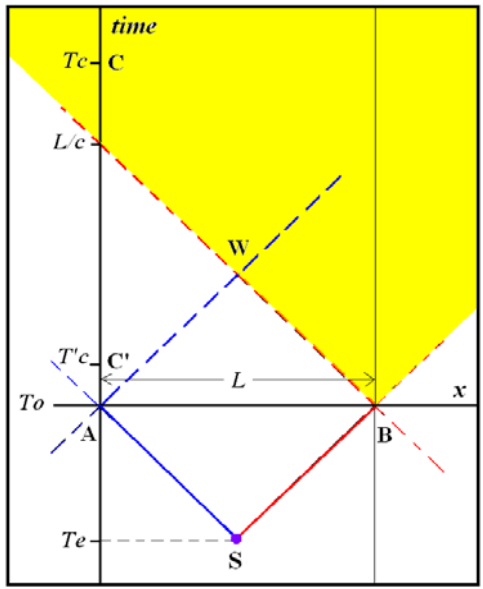

Figure 1: Space-time diagram of the collapse-locality loophole. Source S emits entangled photons that arrive to stations A and B at time $T_o$. If (unknown) collapse time of the A-photon is $T_c \geq L/c$, detection in station A actually occurs as event C, inside the future light cone of B. Event B can be the collapse of the corresponding photon (*essential* loophole), or the definition of the observation setting (*extended* loophole). Events C and B are not space-like separated, so that violation of BI is possible by propagation of information between stations at speed $\leq c$. If the collapse time is $T'c < L/c$ instead, then observations' events C' and B are space-like separated, and BI should *not* be violated.

The core of the collapse-locality loophole is the lack of knowledge on the value of the collapse time $T_c$. If $T_c$ is long enough (collapses occur at C and B), then the observed violation of BI can be explained within LR. If $T_c$ is short enough instead (collapses occur at C' and B), then BI should not be violated. In this case, observing the violation of BI closes the loophole. All attempts to test the loophole [20-27] seek to enforce the collapse in a time shorter than $L/c$, where $L$ is the distance between A and B, and to test the violation of BI. As collapse times are usually predicted to be relatively long, $L$ is large and experiments become difficult. To the best of our knowledge, only one of the proposed experiments, testing the Diósi-Penrose model of gravitational collapse, was actually performed, and required $L = 18$ Km [27]. Here we follow a different approach, based on measuring times of detection rather than violation of BI.

**Experiment.**

It is sketched in Figure 2. Here we present the information needed to understand the results; technical details to allow reproducing the observations are described in the Supplementary Material section.

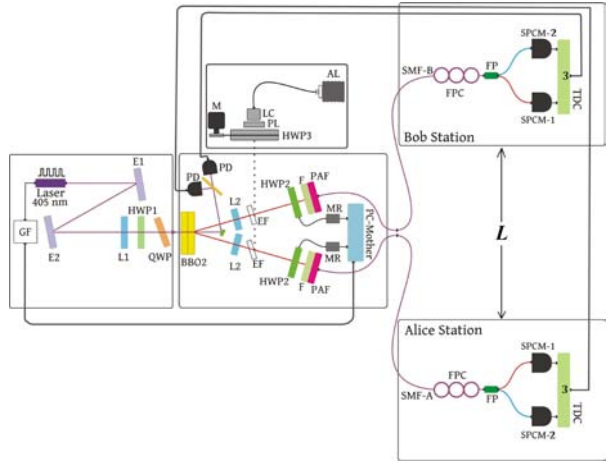

Figure 2: Sketch of the setup. It is a pulsed Bell's experiment with time stamped record of data. GF: function generator that pulses the laser; BBO2: crossed BBO-I crystals (source of entangled states); TDC: time-to-digital converters, they record time values of photons' detections in channels T1 and T2, and trigger signals from pump pulses in T3; PD: photodiodes taking the pulses' samples; SMF-A and B: single-mode fiber coils; FP: fiber polarization analyzer; SPCM: photon counting modules. The stations can be placed at different distances in straight line leaving all other experimental parameters identical; this is a unique feature of this setup. In this paper, observations are made at $L$= 1m and 24m.

A diode laser at 405 nm is pulsed at 500 kHz emitting fairly square shaped pump pulses of duration 500ns, and is focused into two crossed BBO-I crystals in a typical configuration. Photons entangled in polarization in the fully symmetrical Bell's state $|\varphi^+\rangle_{AB} = (1/\sqrt{2}) \cdot \{|x_A\rangle \cdot |x_B\rangle + |y_A\rangle \cdot |y_B\rangle\}$ at 810nm are produced, and propagate through 27m (each) of single-mode optical fibers up to two stations (Alice and Bob) mounted on wheeled tables, which can be placed at adjustable distance $L$ in straight line. This is a unique feature of our setup. In this report, we adjust $L$ to two cases: $L$=1m and $L$=24m.

Each station is identically equipped: a fiber polarizer (FP) separates the beams to silicon avalanche photodiodes (SPCM). The relatively low contrast of these polarizers (<0.98 according to specs) limits the observable value of $S_{CHSH}$ to $\leq 2.77$. We measure $S_{CHSH}$= 2.73 ± 0.07 when $L$= 24m and 2.62 ± 0.07 when $L$= 1m. Time to digital converters (TDC) receive TTL signals from the SPCMs in their input gates #1 and #2, and record the corresponding time values.

Near the source, samples of the pump pulses are sent to two photodiodes (PD). The resulting ("trigger") signals propagate to the stations via 38m (each) of coaxial 50 Ω cables. They are received in input gates #3 of the TDCs. Pulse frequency is modulated by a function generator (GF) according to a pre-established program stored in the "PC-Mother" computer to define pulse numbering and to achieve logical synchronization [29,30]. This feature determines delay values and coincidences unambiguously. Time resolution of the TDCs is in the ps range, but in practice it is limited to ≈2ns because of SPCMs' jitter. Raw data in each station have then the form of three files of time values: one for the trigger (T3), and one for each single-photon detector (T1 and T2). Only one of ≈170 pump pulses produces a detected photon in the average (this figure is necessary in the pulsed regime to keep low the number of accidental coincidences [31]), so that file T3 is much larger than files T1 and T2. These files are then processed: time series of coincidences between the stations (i.e.: T1A-T1B, T1A-T2B, T2A-T1B, T2A-T2B) are obtained using data in T3 to find the correct delays. Single detections are discarded.

Each measuring session is divided in *experiments*, each experiment is a set of 34 *runs*, each run means 30 s of continuous observation in real time. Settings (or bases of observation) are changed for each run. In a typical session four experiments are completed and ≈4.1×10$^6$ coincidences are collected. Settings are mechanically adjusted to scan a wide range of angle values to test the BI. This is a slow variation, so that settings are known across the whole setup well in advance to the arrival of entangled photons to the stations. For this reason, our setup is able to test the *essential* (= detected photons) but not the *extended* (= defined settings) collapse-locality loophole.

Starting from emission at time $Te$ (see Fig.1) the time value of photon detection $Tphot$ stored at the TDCs is given by:

$$Tphot = Te + Tp^{(phot)} + Tc + Tr^{(phot)} + Trec \quad (1)$$

where $Tp^{(phot)}$ is the propagation time of the photons from the source to the stations through the optical fibers, $Tc$ is the collapse time of the quantum state, $Tr^{(phot)}$ is the response time of the SPCM (including cable propagation to the TDC) and $Trec$ is the time required by the TDC to record the time value in its memory. In the same way, the time value of trigger signal detection $Tpulse$ is given by:

$$Tpulse = Te + Tr^{(pulse)} + Tp^{(pulse)} + Trec \quad (2)$$

where $Tr^{(pulse)}$ is the response time of photodiodes, and $Tp^{(pulse)}$ is the propagation time of the trigger signals from the source to the stations through the BNC cables. These are classical signals; hence they are not delayed by a quantum state's collapse time. The difference between eqs.1 and 2 allows calculating $Tc$. Let estimate the involved times now.

The 405 nm laser beam propagates 2.07m (measured) from the BBO crystals to the photodiodes through the air (≈7 ns). The photodiodes (model Thorlabs PDA-100 A2) use an amplifier to deliver a clear signal, for the near-UV radiation is far from the efficiency peak of silicon detectors. They have typical $Tr^{(pulse)}$ = 32 ns (according to specs). Electrical signals propagate through 38m of coaxial cable in 175 ns (measured) to input gates #3 in the TDCs. Hence, signal speed in this type of cable is 4.6 ns/m and $Tpulse ≈ 214$ ns + $Te$ + $Trec$.

The 810 nm entangled radiation propagates 1.06m (measured) through the air from the BBO crystals to the fiberports (≈3.5 ns). Then there are 27m of optical fiber (refraction index ≈1.5) until the SPCMs, which have a response time $Tr^{(phot)}$ = 0.5 ns (according to specs). Coaxial cables (same type as the 38m ones) 2.1m length carry TTL electrical signals from the SPCMs to the corresponding #1 and #2 gates in the TDCs. Hence, $Tphot ≈ 149$ ns + $Te$ + $Trec$. These numbers mean that photons arrive to the TDCs before the "trigger" signals. This is not a problem because the TDCs record input values continuously (as opposed to old oscilloscopes, which started to record information only *after* the trigger's arrival).

**Results.**

As $Trec$ is the same in the three input channels of the TDCs, the difference between the recorded time values of photons and trigger (classical) signals is:

$$Tdif = Tpulse – Tphot ≈ 65 \text{ ns} – Tc \quad (3)$$

Pulses have sharp rising slope; so that the starting time is well determined within 4 ns (see Figure 3 below and the Supplementary Material section). The observed values of $Tdif$, averaged over 136 runs in each case, are displayed in Table 1; dispersions are indicated.

| ⟨$Tdif$⟩ [ns] | $L$=1m | $L$=24m |
|---|---|---|
| T3A – T1A | 56.5 ± 2 | 55 ± 0.2 |
| T3A – T2A | 65.3 ± 1.6 | 64 ± 0.2 |
| T3B – T1B | 60 ± 0.2 | 59 ± 0.1 |
| T3B – T2B | 60.4 ± 2 | 59 ± 0.2 |
| S$_{CHSH}$ | **2.62 ± 0.07** | **2.73 ± 0.07** |

Table 1: Measured differences of detection times between classical signals (T3) and detected photons (T1, T2) averaged over 136 runs for each value of $L$, dispersion are indicated. The S$_{CHSH}$ values measured in each case are also displayed.

In the $L$=1m case, Alice and Bob are so close that, for the time resolution involved, detections are not space-like separated. Unsurprisingly, BI are violated (see bottom row in the table). From eq.3 and Table 1 for $L$=1m, taking the most favorable value:

$$Tc ≤ 65-56.5 = 8.5 \text{ ns} \quad (4)$$

Note this bound is determined from direct observations, and is valid regardless the underlying mechanism of dynamical collapse. It is shorter than predicted by most proposed theories (yet, see a caveat at the Summary).

In the $L$=24m case A and B are 80 ns apart, which is a time much longer than the upper bound of $Tc$ in eq.4. In consequence, detections in this case occur as space-like separated events, like C' and B in Fig.1. All experimental features (threshold values in the TDCs, individual detectors used, optical fibers, etc.) are identical than in the $L$=1m case, the only difference is the distance in straight line between the stations. Therefore, if the essential collapse-locality loophole is valid, there are only two possible observable outcomes: *i)* BI are not violated, or *ii)* detections are delayed until the information from the collapse's outcome reach the other station, i.e., 80-8.5 ≈ 72 ns. As seen in the Table, BI are amply violated and the time value photons are detected (relative to the position of the trigger) when *L*=24m is the same as when *L*=1m within the experimental uncertainty (4 ns); the shift ≈72 ns is not observed. None of the two outcomes consistent with the essential collapse-locality loophole is observed, so we consider the observations close this loophole. Be aware that nothing can be said about the *extended* loophole, for (as already stated) the settings are changed too slowly.

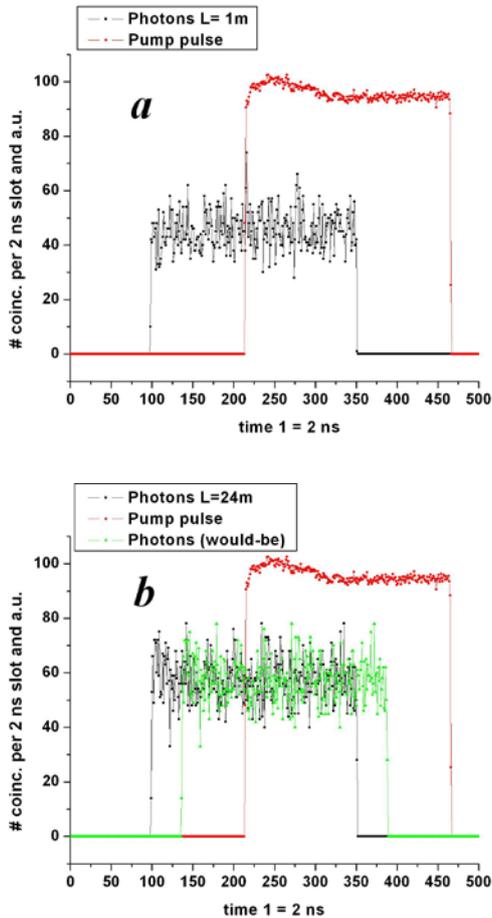

Figure 3: Number of coincidences "1,1" per 2ns time slot recorded during a single run with angle settings A=π/4, B= π/8, the pump pulse (in red) is fixed at arbitrary time = 214 (note the 2 ns scale); (*a*) *L*=1m; (*b*) *L*=24m, photons at (*b*) start 1.5 ns closer to the pulse than in (*a*), what is not noticeable at this scale; the would-be position of the photons in (*b*) if the second alternative for the loophole were true (delayed 72 ns, in green) is also plotted.

Figure 3 is an illustration of experimental results; it displays the time position, with respect to the pump pulse, of "1,1" coincident photons' detections. Number of coincidences in each 2 ns time slot recorded in a single run for setting A=π/4, B=π/8 are plotted. The pump pulse profile is extracted from a file of single detections and scaled in arbitrary vertical units. Note there is no observable difference of the relative time positions between the *L*=1m and *L*=24m cases (actually, it is 1.5 ns). In (*b*) it is also plotted the position of the photons shifted 72 ns, to illustrate how it would look like if the second alternative (to save the essential collapse locality loophole) was true.

Besides the essential and extended versions of the loophole, there is a third possibility [18], which is that the collapses occur when the information from both stations is gathered in a point in the future of both arrival events, i.e., point W in Fig.1, or later. In our setup, this implies that the difference between the arrival of the classical signal in #3 and the other two channels should change $L/2c$ ≈ 40 ns between the *L*=1m and *L*=24m cases. This is not observed either, what disproves this possibility too.

**Summary.**

By carefully considering time values of detection instead of enforcing collapses, the essential cavity-loophole is experimentally closed, independently of the underlying theory of dynamical collapse. This is achieved thanks to the special features of our setup, that is: *i)* the existence and recording of classical signals free of quantum collapse delay; this allows measuring times of detection with respect to a well established start value, and *ii)* the possibility of measuring with an identical setup at different distances between the stations of observation.

Caveat: it is still possible that $T_c$ is near to a multiple of the pump pulses' time separation, i.e., that detected photons do not correspond to the nearest pump pulse, but to a pulse that was emitted one or more periods (of ≈2 μs) before. Yet, the pulse repetition rate slightly changes during the runs, and the reported experiment was repeated with different repetition rates obtaining the same results. Therefore, defenders of the tested loophole should suppose the system "knows" in advance the repetition rate that will be used, and conspires accordingly to hide the actual value of $T_c$. Strictly speaking this is possible, for the modulation/repetition rate is stored in the memory of the "Mother" PC long time before the experiment starts. As said in the Introduction, what experiments can achieve is not a perfect refutation, but a reduction of the space allowed to the loopholes down to the point that they become too conspiratorial to be acceptable. Whether that point has been reached in this case is left to the Reader's opinion.

Finally, the bound in eq.4 is obtained for this particular setup. Nevertheless, it is a reference value for $T_c$ that may be useful to the theories of dynamical quantum state collapse in the future.

**Acknowledgments.**

This work received support from grant PIP 00484-22 from CONICET (Argentina).

**References.**

**SUPPLEMENTARY MATERIAL** for: "Test of the essential collapse-locality loophole"; Mónica Agüero, Juliana Bourdieu, Alejandro Hnilo, Marcelo Kovalsky and Myriam Nonaka.

**Experimental details.**
*Description of the setup.*

Figure SM1 is a detailed diagram of the experimental setup. Biphotons at 810 nm in the fully symmetrical entangled Bell state $|\varphi^+\rangle$ are produced in two crossed (1mm long each) BBO-I crystals, pumped by a pulsed diode laser at 405 nm. The average CW laser power is 40 mW. Coherence length is measured 40 mm at 100 kHz pulse repetition rate and 10% duty cycle. This means $\Delta\omega_p \approx 5\times10^{10}$ s$^{-1}$, which is much smaller than the bandwidth of the spontaneous photon down conversion process in the crystals, and also than the filters' bandwidth $\Delta\omega_f \approx 3\times10^{13}$ s$^{-1}$. Coherence length is observed to increase with duty cycle (in CW operation it is 20 m according to specs). This laser is able to emit fairly square pulses of adjustable duration and repetition rate. Pulse shape deteriorates if the rate is higher than 1MHz or pulse duration is shorter than 200 ns. Duty cycles as low as 5% have been used with satisfactory results.

After producing the pair of entangled photons, the pump beam is sent to a 50-50 beam splitter. The output beams illuminate two fast photodiodes (model Thorlabs PDA-100 A2), which produce electrical signals defining the start of each pump pulse. These signals are sent to the stations "Alice" and "Bob" through 50Ω coaxial cables 38 m length each. They are checked to have negligible distortion. Two photodiodes are used, instead of just one with a "T" BNC, because of spurious echoes in the long cables. The classical signals indicating the start of each pump pulse enter the #3 input channels of the time-to-digital converters (TDCs, Id Quantique Id-900), one in each station. These are the largest files, because most pulses are "empty". Only ≈0.6% of the pulses produce detected photons. This is necessary to keep low the number of accidental coincidences in the pulsed regime. In a typical run, tens of millions of trigger signals must be recorded correctly by both TDCs, what is challenging.

In order to keep tracking of pulse numbering with independent clocks (which unavoidably drift away), the repetition rate is switched or modulated following a previously established protocol, in order to establish a "logical" synchronization between the clocks in the two TDCs. Therefore, the pulsed regime refreshes the synchronization between the distant clocks with each pulse; the frequency modulation allows reliable pulse numbering and immediately determines the correct delays between the lists of photons' detections. There is no need of convoluting the time lists and counting coincidences for each possible value of delay. This is an important practical advantage, and provides unambiguous and reliable results.

The entangled beams propagate through single-mode optical fibers (S630-HP Nufern) 27 m long each, which are extended from the source to the stations. The fibers are inserted into flexible stainless steel tubes (12 mm inner diameter, 16 mm external), which traverse the lab's walls through drilled holes to the adjacent corridor and are placed over cable trays until reaching the stations (see Figure SM1, DOWN). Both stations are identically equipped. Optics and electronics are mounted on small wheeled optical tables. Optics is placed inside a black box that protects it from spurious light and dust. The stainless steel tubes reach into these boxes. When measuring at short distance in straight line *L*, the stations are moved inside the lab and the tubes are bent to re-enter the lab through its door. Excepting for the distance in straight line between the stations, observations in the corridor or within the lab have identical values of all experimental parameters.

In each station, "bat-ears" are used to compensate birefringence distortion in the fibers. Polarization is observed with two-exit fiber optic analyzers (Thorlabs PFC-780SM-FC). An auxiliary laser beam at 808 nm is inserted into the fibers at the source through removable mirrors EF (see Fig.SM1). The bat-ears are adjusted to maximize contrast between the two output gates of the fiber polarizers. For this action the SPCM are replaced by standard photodiodes. In order to facilitate measuring the contrast, a half waveplate (HWP3) in a fast (about 100 Hz) rotating mount is inserted between the auxiliary laser and the fiberports optics. Transmission from the input of the focusing fiberports optics (PAF, see Fig.SM1) to the detectors is measured 83% for Alice and 82% for Bob.

The beams leaving the two outputs of the fiber polarizers are sent to single photon counting modules (SPCM, AQR-13 and AQRH-13, from Perkin-Elmer-Pacer-Excelitas). These modules emit one TTL signal for each detected photon. Efficiency according to specs is ≈65% at the operation wavelength. The TTL signals are sent to input channels #1 (outcome "1") and #2 (outcome "0") of the TDCs in each station. Detections' time values are stored. The TDCs have 10 ps nominal time resolution, but accuracy is reduced to 2 ns because of SPCM's jitter. One PC in each station controls the duration of the observation run, the opening and closing of files and their naming and saving, following the instructions sent via wifi by a "Mother" PC placed near the source.

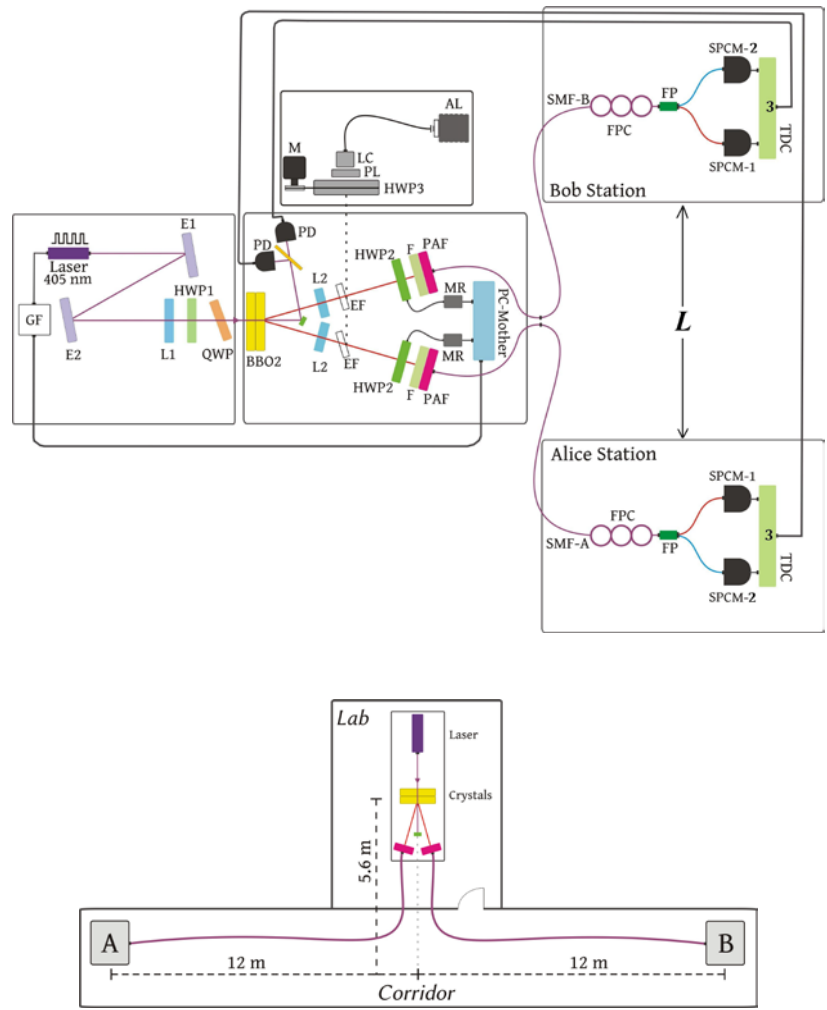

Figure SM1: UP: Setup's diagram. GF: function generator that modulates the laser repetition rate according to a previously established program; L1,L2: $f$= 300 mm lenses; E1, E2: HR plane mirrors at 405 nm; HWP1 and QWP: half and quarter waveplates at 405 nm; BBO2: crossed BBO-I crystals (source of entangled states); PD: fast photodiodes, they send trigger signals of pulses' emissions to the TDCs via coaxial 50 Ω cables; HWP2, HWP3: half-waveplates at 810 nm; F: Interferential filters at 810 nm, Δλ=10 nm, 90% transmission; EF: auxiliary, removable HR plane mirrors in flip-flop mountings; AL: auxiliary CW laser diode at 810 nm coupled to a multi-mode fiber; LC: collimating optics; PL: linear polarizer; M: motor that rotates HWP3; MR: servo motor controllers of HWP2; PAF: fiberports f=7.5mm; SMF-A and B: single-mode fiber coils, 27 m total length each; FPC: birefringence compensator ("bat-ears"); FP: fiber polarization analyzer; SPCM: photon counting module; TDC: time-to-digital converters. The stations can be placed at different distances in straight line, what is a unique feature of this setup. DOWN: Scheme of the stations' and source's positions when $L$= 24m.

"Mother" directly controls the function generator that pulses the laser (including the switching or modulation of the repetition rate, following a previously specified program) and the servo motors that adjust the settings angles. She also controls remotely wifi through a TCP/IP communication via a local network, the "sons" PCs in each station (Alice and Bob) to open, name and close the data files recorded in each TDC. Raw data are saved in .bin format. For a single 30 s run they require typically 300 kbit for each file of photons' detections times, and 120 Mbit for the file of pulses' detection times. After being processed and summed up, data of a whole session (typically ≈200 runs) require less than 20 Mbit in .dat format. We are eager to share our raw and/or processed data upon reasonable request.

The coaxial cables carrying the pulses' start signal are 38 m long each, but the fibers are only 27 m long each. This means that "trigger" signals arrive to the TDCs later than "signal" photons. This is not a problem, for the TDCs record data continuously in all the input channels. The delay is observed to be constant (≈65 ns) and is taken into account for data processing. Photons' detection times are positioned in reference to the trigger. Note that, in each station, time values in the three channels are measured by a single clock. Synchronization between the clocks in each station is achieved through the trigger pulses arriving to channels #3, and slight modulation of the pulses' frequency, as explained.

*Procedure of data recording.*

We call one "run" an interrupted period of recording data. Data are saved in the PCs at the end of each run. Once the controlling program is started, the setup is able to perform an arbitrary number of successive runs with different settings (which are previously specified) without the operators' assistance.

Typically, each run records data during 30s of real time. Recording runs are gathered in sets named "experiments", which accumulate the results of 34 runs, scanning a complete set of angle values {α,β}. Because of idle periods of time inserted to give time the PCs to download from the TDCs and save the data files reliably, each "experiment" lasts about one hour. Typical curves of coincidences obtained in one "experiment" are shown in Figure SM2.

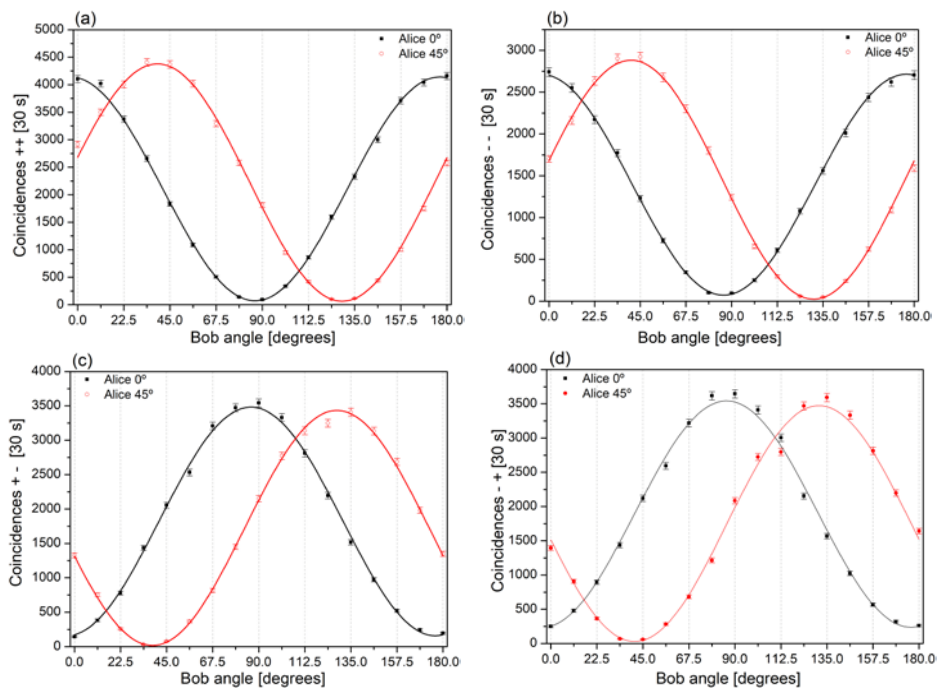

Figure SM2: Illustration of the curves recorded in one "experiment"; (a) Total number of coincidences as a function of the setting angles for "+,+" coincidences (i.e., coincidences between the detectors "+" in each station), (b) for "-,-", (c) for "+,-", (d) for "-,+". In this case, measured $S_{CHSH}$ = 2.75, $L$=24 m.

A first checking experiment is performed, and sets of curves of coincidences as function of angle settings {α,β} are drawn. If the obtained curves are not satisfactory, realignment and/or improved birefringence compensation are performed. The checking experiment is then repeated. If everything is satisfactory instead, several further experiments are carried out until sufficient statistics is accumulated. Typically, four valid experiments are performed during each session.

Photons' detections times (which correspond to collapses of quantum states) in channels #1 and #2 are positioned in reference to the classical trigger signals in channel #3. After summing up data produced by millions of pulses ($\approx 5 \times 10^8$ in a single experiment), plots of Singles and Coincidences as a function of time are obtained with satisfactory statistics. Frequency modulation of the pulses' frequency ensures the numbering of each pumping pulse to be the same in each station. Data processing shows that detections occurring during pulses with the same numbering are coincident within 2 ns. There are practically no coincident detections observed outside the pump pulses, which agrees with the following estimation: for a 4 ns coincidence time window and detectors' typical dark count rate of 200 s$^{-1}$, the number of accidental coincidences accumulated during a 30 s run is: $(200 \text{ s}^{-1})^2 \times 4.10^{-9} \text{ s} \times 30 \text{ s} \approx 5 \times 10^{-3}$ in each time slot. Therefore, even if many runs are accumulated ($\approx 200$ during one session of 6 experiments), only rarely one coincidence occurs in a slot outside the pulses.

As an illustration, typical time variations of singles and coincidences for one of the detectors (A+) are displayed in Figure SM3. These curves match the laser pump pulse shape as observed with a fast photodiode. The small fluctuations from the ideal pulse shape do not appear in $S_{CHSH}(t)$, which is remarkably square (see, f.ex.: M.Agüero *et al.*, "Testing transient deviations from quantum mechanics' predictions in spatially spread optical entangled states", *J.Opt.Soc.Am.B* **42**(2), 2025).

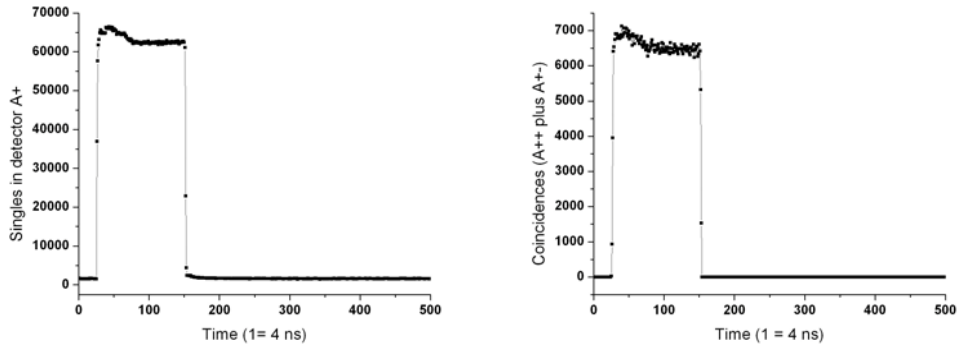

Figure SM3: Time evolution of singles and coincidences in detector "+" in station Alice (A+). Left: single detections, Right: coincidences (A++ plus A+-). Coincidences outside the pump pulses are zero, as expected. The number of singles outside the pump pulses ($\approx 1600$ in each 4 ns slot) is consistent with the measured rate of dark counts of detector A+ (140 s$^{-1}$).

*Birefringence compensation.*

In order to make easier the (otherwise mostly try-and-error) method to compensate birefringence in the fibers with bat-ears, we use an auxiliary laser diode at 810 nm. This is practically the same wavelength of the entangled photons. The laser is fiber coupled (multi-mode). The beam is collimated, polarized and passed through a half wavelength waveplate. This waveplate is placed in a rotating mounting driven by a motor at ≈50 Hz. The result is a polarized beam at 810 nm which plane of polarization rotates at ≈100 Hz. This beam is inserted into the single mode fibers by using high reflection (at 45°) plane mirrors on flip-flop mountings (so that the mirrors can be easily removed from the optical axes, see Fig.SM1).

In each station, two photodiodes are placed at the exits of the fiber polarizer, and their outputs observed in an oscilloscope, see Figure SM4. The first coil in the bat ear (a quarter waveplate) is adjusted to maximize contrast. This means the point (representing the laser field) in the Poincaré sphere to be in the equator. Then the motor is stopped, and the position of the rotating waveplate is manually adjusted to the position that corresponds to one of the output fibers in the fiber polarizer (this position was previously marked on the support of the rotating mounting). Then the half waveplate in the bat ear is adjusted to maximize the signal (which is constant in time now) in one of the photodiodes, and to minimize the signal in the other. Then the waveplate is put into rotation again, and fine adjustments are introduced. The free coil in the bat-ear (which is theoretically unnecessary) is used, in practice, to adjust contrast in the diagonal basis of polarization. The procedure is repeated at the other station.

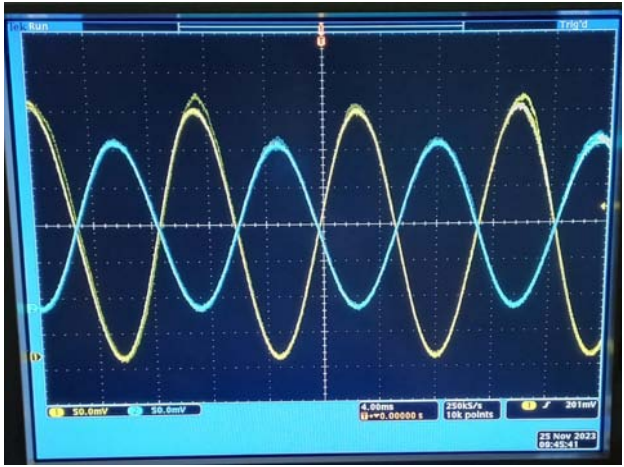 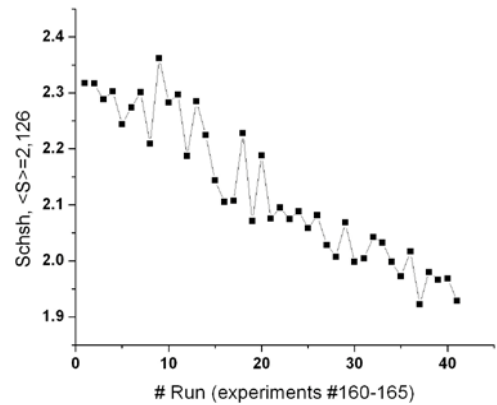

Figure SM4: Left: appearance of the photodiodes' signals when the auxiliary waveplate is rotating, the "bat-ears" are adjusted to maximize contrast of the (opposite phase) sinusoids. Right: example of the decay of $S_{CHSH}$ along the day (about 10 hs) in a condition of deficient birefringence compensation at start.

Thermal and mechanical perturbations affect birefringence during the day. In Fig.SM4, the decay of the value of $S_{CHSH}$ with real time is shown along the day during a period of about 10 hs. This puts a practical limit to the total duration of a measuring session without new adjustments. Careful birefringence compensation using two photodiodes as described above is proven to be important to get starting values good enough to allow recording data during a whole day without needing further adjustments.